\documentclass{blois}

\usepackage{soul}

\def\be{\begin{equation}}
\def\ee{\end{equation}}
\def\bea{\begin{eqnarray}}
\def\eea{\end{eqnarray}}

\begin{document}
\vspace*{4cm}
\title{CONSTRAINING DECAYING DARK MATTER WITH THE
EFFECTIVE FIELD THEORY OF LARGE-SCALE STRUCTURE}

\author{TH\'EO SIMON}

\address{Laboratoire Univers \& Particules de Montpellier (LUPM), CNRS \& Universit\'e de Montpellier (UMR-5299),Place Eug\`ene Bataillon, F-34095 Montpellier Cedex 05, France}

\maketitle

\abstracts{I present the first constraints on decaying cold dark matter (DCDM) models thanks to the effective field theory of large-scale structure (EFTofLSS) applied to BOSS-DR12 data. I consider two phenomenological models of DCDM: i) a model where a fraction $f_{\rm dcdm}$ of cold dark matter (CDM) decays into dark radiation (DR) with a lifetime $\tau$; ii) a model (recently suggested as a potential resolution to the $S_8$ tension) where all the CDM decays with a lifetime $\tau$ into DR and a massive warm dark matter (WDM) particle, with a fraction $\varepsilon$ of the CDM rest mass energy transferred to the DR. I discuss the implications of the EFTofLSS constraints for the DCDM model suggested to resolve the $S_8$ tension.}

\section{Introduction}

The core cosmological model, known as the $\Lambda$ cold dark matter ($\Lambda$CDM) model, delivers an exceptional explanation for a broad variety of early and late Universe data. However, as the accuracy of measurements has increased over the past few years, some intriguing discrepancies have emerged within this model:  for instance, the ``$S_8$ tension''~\cite{Abdalla:2022yfr} corresponds to a mismatch of the value of the amplitude of the local matter fluctuations (typically parameterized as $S_8$) between its prediction by the $\Lambda$CDM model from the CMB data~\cite{Planck:2018vyg,ACT:2020gnv} on the one hand, and its direct and local determination on the other hand~\cite{KiDS:2020suj,DES:2021wwk,HSC:2018mrq,Amon:2022ycy} (the latter being weaker than the former). The $S_8$ parameter is define as $S_8 = \sigma_8 \sqrt{\Omega_m/0.3}$, where $\Omega_m$ is the current total matter abundance, and $\sigma_8$ is the root mean square of matter fluctuations on a $8 \ h^{-1}$Mpc scale (with $h = H_0/(100 \ \textrm{km/s/Mpc})$).
This tension could be the first indication of new features in the dark components which would cause a decrease in the power spectrum at scales 
$k \sim 0.1 - 1$ Mpc/$h$.
We present here~\cite{Simon:2022ftd} two decaying cold dark matter models that have the ability to lead to such a suppression.~\cite{FrancoAbellan:2020xnr,FrancoAbellan:2021sxk,Audren:2014bca,Enqvist:2015ara,Poulin:2016nat,Aoyama:2014tga,Alvi:2022aam,Holm:2022eqq,Fuss:2022zyt,Holm:2022kkd,Nygaard:2020sow,Holm:2022kkd,Bucko:2022kss}

Here, we go beyond previous works by making use of the effective field theory of large-scale structure (EFTofLSS)~\cite{Carrasco:2012cv,Baumann:2010tm,Porto:2013qua,Senatore:2014eva} to describe the mildly non-linear regime of the galaxy clustering power spectrum and derive improved constraints thanks to the EFTofLSS applied to BOSS data~\cite{BOSS:2016wmc,Reid:2015gra} (EFTofBOSS). Despite some precautions to be taken in the interpretation of the results,~\cite{Simon:2022lde} the EFTofBOSS data have been shown to allow for the determination of the $\Lambda$CDM parameters~\cite{DAmico:2019fhj,Ivanov:2019pdj,Colas:2019ret,Philcox:2020vvt,Chen:2021wdi,Zhang:2021yna,Chen:2022jzq,Simon:2022lde,Simon:2022csv,Smith:2022iax} at a precision higher than that from conventional BAO and redshift space distortions (denoted as ``BOSS BAO/$f\sigma_8$''), as well as to provide interesting constraints on models beyond $\Lambda$CDM.~\cite{Colas:2019ret,Ivanov:2019hqk,Kumar:2022vee,DAmico:2020kxu,DAmico:2020tty,Carrilho:2022mon,Chudaykin:2020ghx,Glanville:2022xes,DAmico:2020ods,Ivanov:2020ril,Niedermann:2020qbw,Simon:2022adh,Simon:2022ftd,Rubira:2022xhb,Nunes:2022bhn,Lague:2021frh}

\section{Constraints on the DCDM $\to$ DR model}

\subsection{Presentation of the model}

In the first model we consider,~\cite{Poulin:2016nat} the CDM sector is  partially composed of an unstable particle (denoted as DCDM) that decays into a non-interacting relativistic particle (denoted as DR). The rest of the DM is considered stable and we refer to it as the standard CDM. 
In addition to the standard six $\Lambda$CDM parameters, there are two free parameters describing the lifetime of DCDM $\tau$ (or equivalently the decay width $\Gamma = \tau^{-1}$), as well as the fraction of DCDM to total dark matter at the initial time $a_{\rm ini}\to 0$: $f_{\rm dcdm} \equiv\omega_{\rm dcdm}(a_{\rm ini}) / \omega_{\rm tot,dm}(a_{\rm ini})$,
with  $\omega_{\rm tot, dm}\!\equiv\! \omega_{\rm dcdm} + \omega_{\rm cdm}$. With these definitions, in the limit of large $\tau$ and/or small $f_{\rm dcdm}$, one recovers the $\Lambda$CDM model. The evolution of the homogeneous energy densities of the DDM and DR is given by~\cite{Poulin:2016nat}:
\begin{eqnarray}
    \dot{\bar{\rho}}_{\rm dcdm} + 3 \mathcal{H}\bar{\rho}_{\rm dcdm} = -a\Gamma \bar{\rho}_{\rm dcdm} ~~~;~~~ \dot{\bar{\rho}}_{\rm dr} + 4 \mathcal{H}\bar{\rho}_{\rm dr} = a\Gamma \bar{\rho}_{\rm dcdm},
\end{eqnarray}
where $\mathcal{H}$ is the conformal Hubble parameter.

To describe the evolution of the linearly perturbed universe, we consider the usual synchronous gauge, where the frame co-moving with the DCDM (and CDM) fluid. Since we consider a homogeneous and isotropic decay, the energy density perturbation of the DCDM component, $\delta_{\rm dcdm} \equiv \rho_{\rm {dcdm}}/\bar{\rho}_{\rm {dcdm}}-1$, follows the same evolution as standard CDM. Consequently, the specific effects of the DCDM $\to$ DR model occur at the background level (since the effect of the daughter particles is minor).

\begin{figure*}
    \centering
    \includegraphics[trim=2cm 10.5cm 3cm 1.8cm, clip=true, width=0.87\columnwidth]{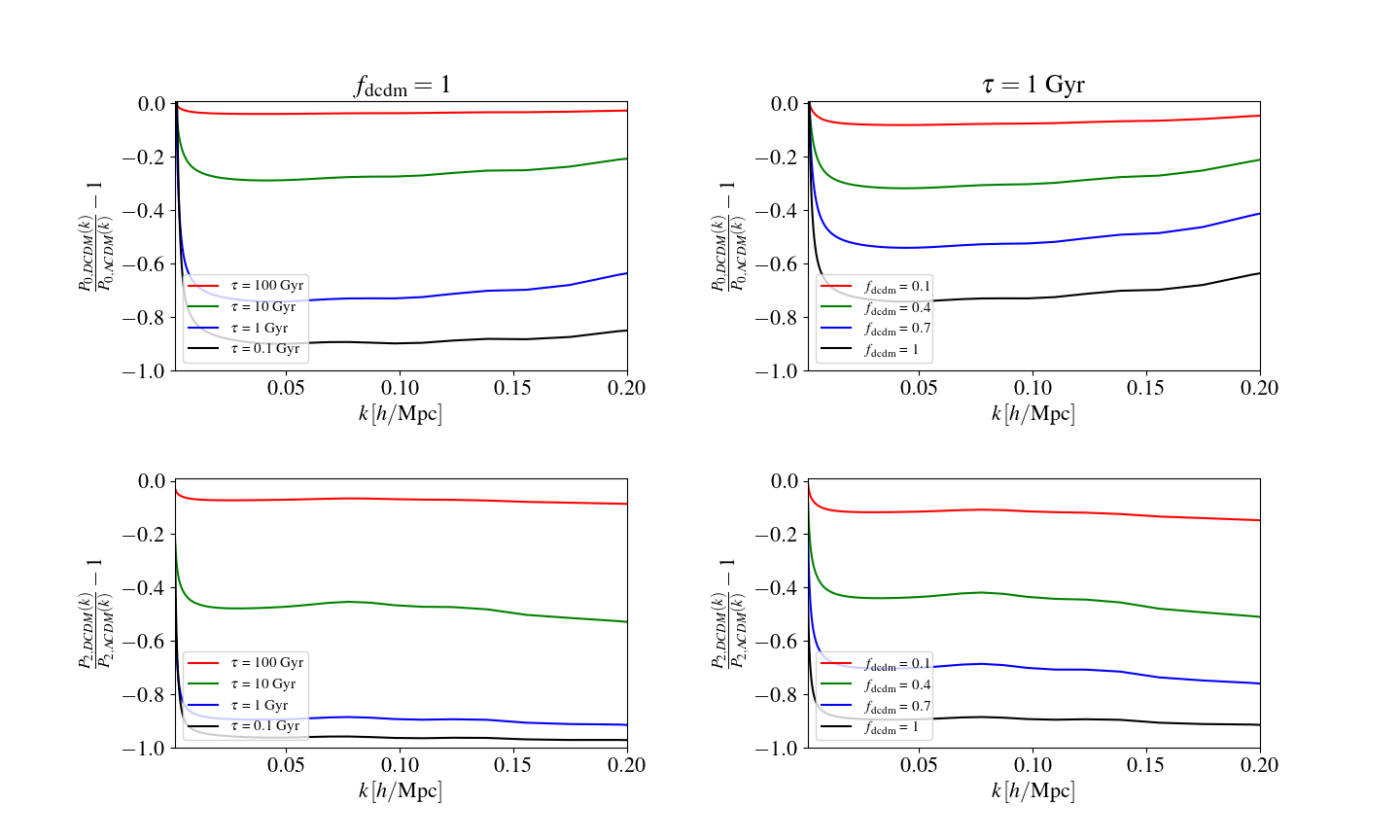}
    \includegraphics[trim=2cm 10.5cm 3cm 1.8cm, clip=true,width=0.87\columnwidth]{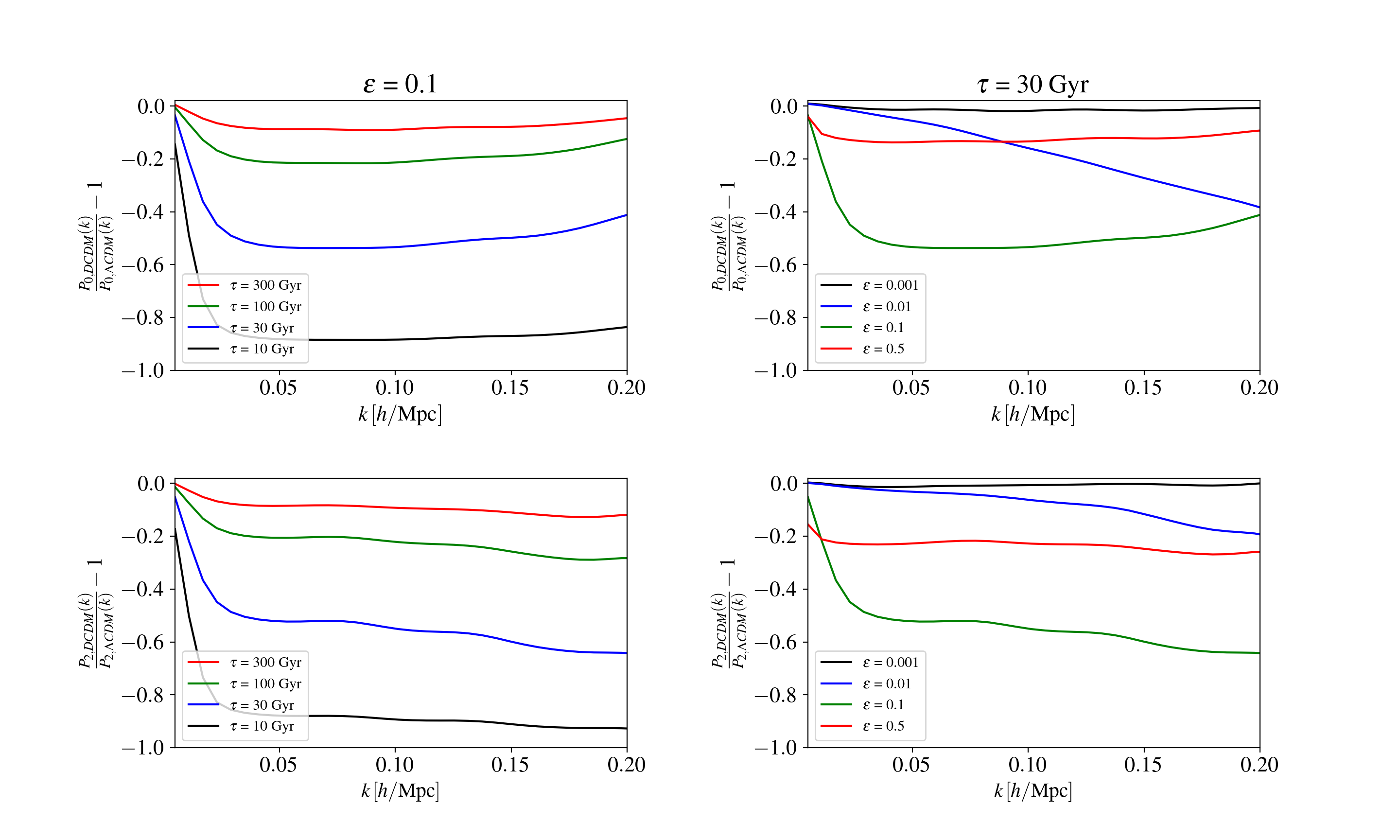}
    \caption{\textit{Upper} - Residuals (with respect to the $\Lambda$CDM model at $z=0$) of the monopole of the DCDM $\to$ DR galaxy power spectrum for several values of $f_{\rm dcdm}$ and $\tau$. \textit{Lower} - The same applies to the DCDM $\to$ WDM+DR model with several values of $\tau$ and $\varepsilon$.}
    \label{fig:pybird_dcdm_dr}
\end{figure*}

In the upper panel of Fig.~\ref{fig:pybird_dcdm_dr}, we represent the residuals of the monopole of the galaxy power spectrum for several values of $f_{\rm dcdm}$ and $\tau$ to isolate their cosmological effects: as expected, deviations with respect to $\Lambda$CDM increases as  $\tau$ decreases and/or $f_{\rm dcdm}$ increases. This deviation takes the appearance of a power suppression due to two main reasons.~\cite{Poulin:2016nat} First, the decay of DCDM decreases the duration of the matter dominated era (and at fix $h$, a smaller $\Omega_m$/larger $\Omega_{\Lambda}$), implying a shift of the power spectrum towards large scales, {\it i.e.} towards small wavenumbers. 
Second, DCDM models involve a larger ratio of $\omega_{\rm b}/\omega_{\rm cdm}$ compared to the $\Lambda$CDM model due to the decay.
Both effects manifests as a strong suppression of the small-scale power spectrum, and the latter effect leads to an additional modulation of the BAO amplitude visible as wiggles in the upper panel of Fig.~\ref{fig:pybird_dcdm_dr}.

\subsection{EFTofLSS Constraints on the DCDM $\to$ DR model}

\begin{figure*}
    \centering
    \includegraphics[width=0.49\columnwidth]{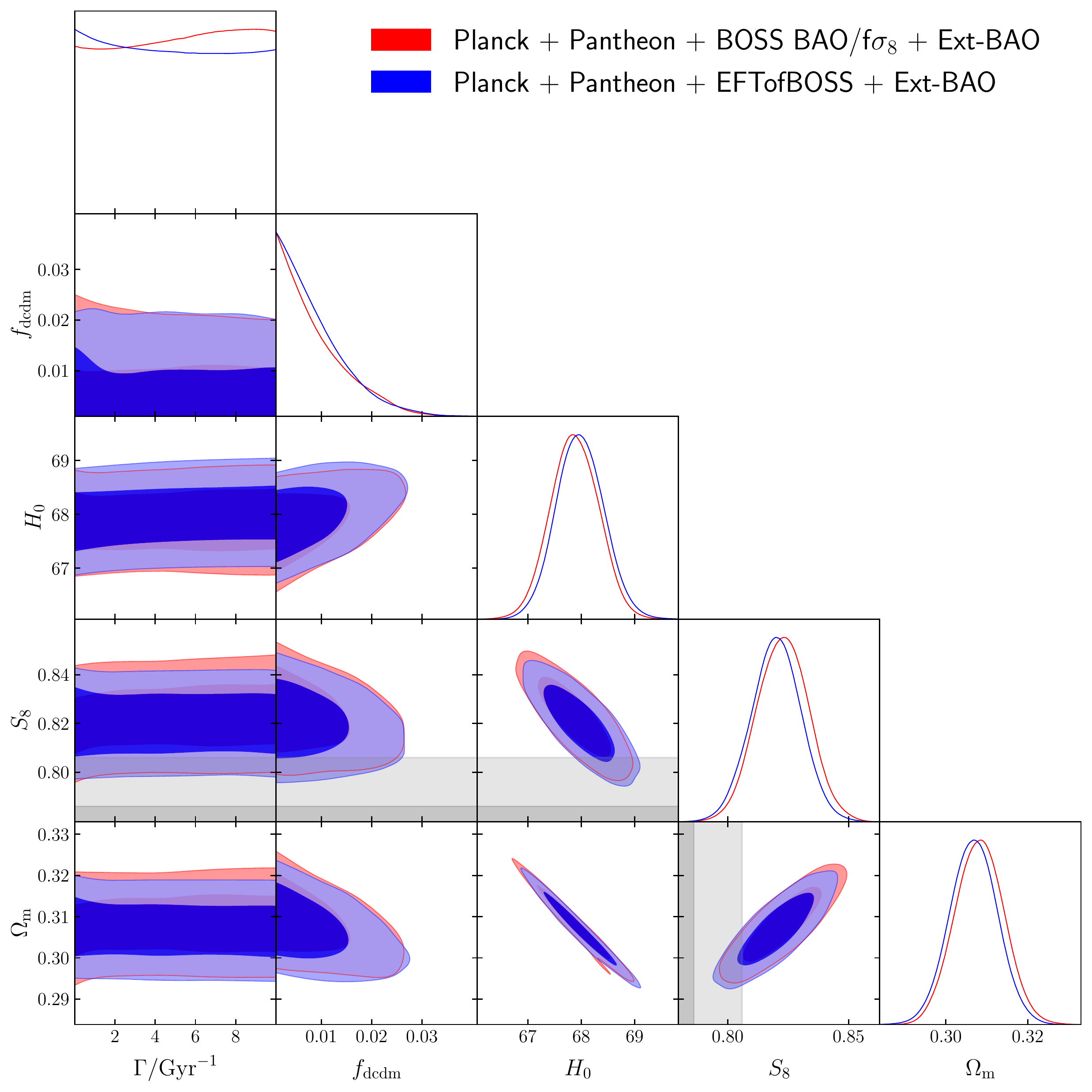}
    \includegraphics[width=0.49\columnwidth]{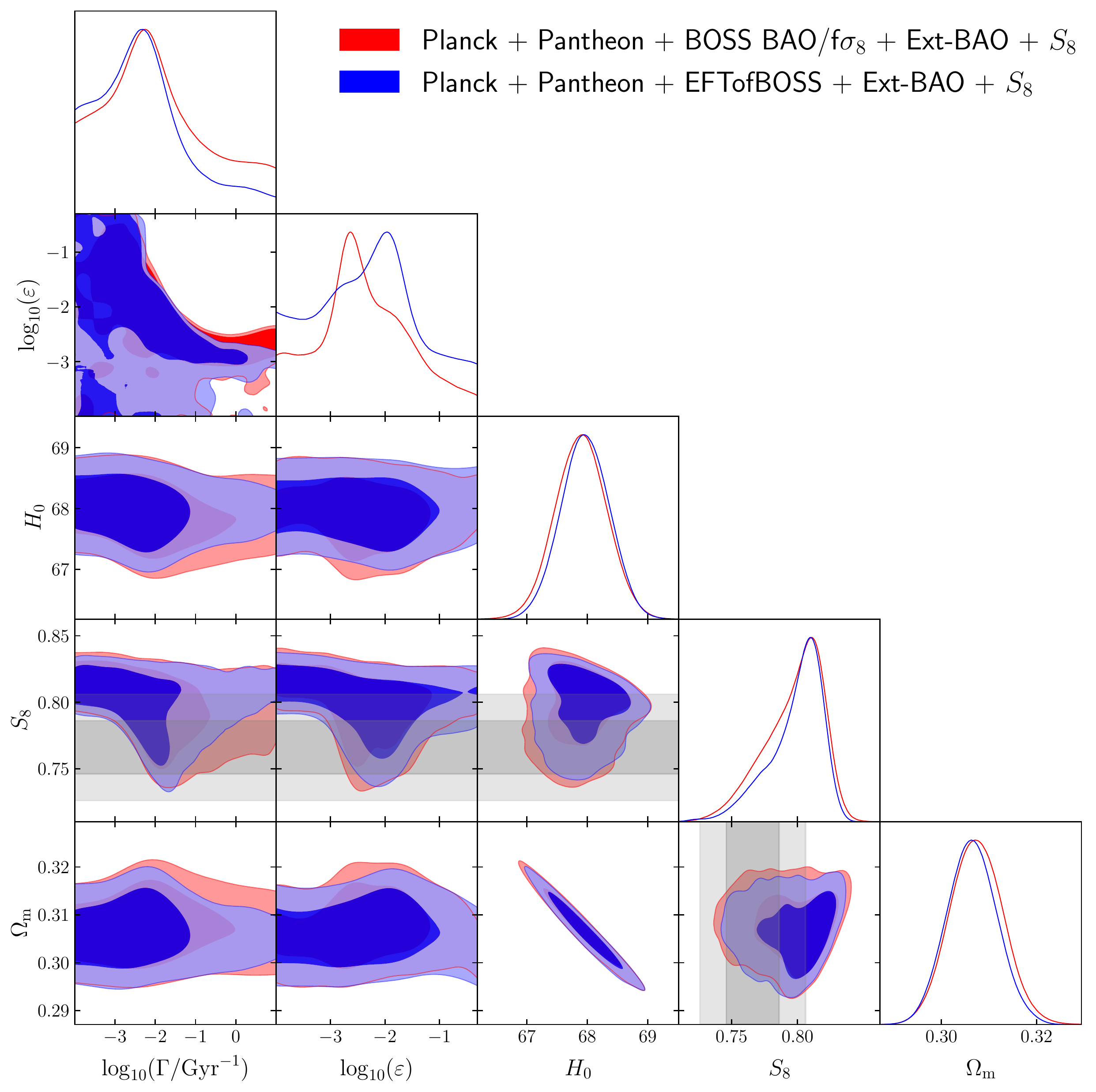}
    \caption{{\it Left} - 2D posterior distributions of the DCDM $\to$ DR model with and without the EFTofBOSS dataset. The gray shaded bands refer to the joint $S_8$ measurement from KiDS-1000 + BOSS + 2dFLens. {\it Right} - 2D posterior distributions of the DCDM $\to$ WDM+DR model with and without the EFTofBOSS dataset.}
    \label{fig:constraints}
\end{figure*}

In the left panel of Fig.~\ref{fig:constraints}, we display the 1D and 2D posteriors of the reconstructed parameters for the DCDM $\to$ DR model with and without the EFTofBOSS dataset. When we do not take into account the data from EFTofBOSS, we include the standard RSD information, denoted as ``BOSS BAO/$f\sigma_8$''. In these analysis, we always take into account data from {\it Planck}~\cite{Planck:2018vyg} (namely the CMB power spectra as well as the gravitational lensing potential), Pantheon~\cite{Scolnic:2017caz} and Ext-BAO~\cite{Beutler:2011hx,Ross:2014qpa,Alam:2016hwk,Agathe:2019vsu} (which corresponds to measurements using the BAO information without EFT treatment). 

One can see that the inclusion of the EFTofBOSS data does not improve the constraints on this model with respect to the standard RSD information. However, while $\tau$ is unconstrained, we obtain an interesting 95 \% C.L constraint on the DCDM fraction : $f_{\rm dcdm} < 0.0216$. We perform a second analysis~\cite{Simon:2022ftd} where we set $f_{\rm dcdm} = 1$ in order to derive a lower constraint on $\tau$. We obtain, at 95 \% C.L., $\tau > 249.6$ Gyr.

Finally, we show~\cite{Simon:2022ftd} that when adding a $S_8$ prior from the KIDS-1000 cosmic shear measurement, the $\Delta\chi^2$ with respect to $\Lambda$CDM is still compatible with zero. We conclude (as in past studies) that this model does not resolve the $S_8$ tension. This is because the suppression of the power spectrum (which could allow us to resolve the $S_8$ tension) is mainly due to a background effect, through the decrease of $\Omega_m$. This latter parameter is however well constrained by the different data we have taken into account.

\section{Constraints on the DCDM $\to$ WDM+DR model}

\subsection{Presentation of the model}

We now turn to a DCDM model~\cite{FrancoAbellan:2021sxk} where the entirety of the DM sector is considered unstable ({\it i.e.}, $f_{\rm dcdm}=1$ in the language of the first model), decaying into dark radiation and a massive particle, which act as warm dark matter (WDM). 
As before, we assume the decay products do not interact with the standard model particles.
The DCDM sector is now described by the DCDM lifetime  $\tau$, and the fraction $\varepsilon$ of rest-mass energy carried away by the massless particle  given by $\varepsilon = 1/2 \cdot \left(1- m^2_{\rm wdm} / m^2_{\rm dcdm}\right)$, where $m_{\rm dcdm}$ and $m_{\rm wdm}$ are the mother and daughter particle masses respectively. 
The set of equations describing the evolution of the background energy densities of the dark components reads as follows~\cite{FrancoAbellan:2021sxk}:
\begin{eqnarray}
   &\dot{\bar{\rho}}_{\rm dcdm} + 3 \mathcal{H}\bar{\rho}_{\rm dcdm} = -a\Gamma \bar{\rho}_{\rm dcdm} ~~~;~~~
    \dot{\bar{\rho}}_{\rm wdm} + 3(1+w) \mathcal{H}\bar{\rho}_{\rm wdm} = (1-\varepsilon)a\Gamma \bar{\rho}_{\rm dcdm} ~~~; \\
   & \dot{\bar{\rho}}_{\rm dr} + 4 \mathcal{H}\bar{\rho}_{\rm dr} = \varepsilon\Gamma
a \bar{\rho}_{\rm dcdm},
\end{eqnarray}
where $w=\bar{P}_{\rm wdm}/\bar{\rho}_{\rm wdm}$ is the equation of state of the WDM. 
In the limit of large $\tau$ or small $\varepsilon$, one recovers the $\Lambda$CDM model, while setting $\varepsilon=1/2$ leads to a decay solely into DR.

In the lower panel of Fig.~\ref{fig:pybird_dcdm_dr}, we represent the residuals of the monopole of the galaxy power spectrum for several values of $\varepsilon$ and $\tau$ to isolate their cosmological effects. 
In this model, $\tau$ -- which sets the abundance of the WDM species today -- controls the amplitude of the power suppression, while $\varepsilon$ controls the cutoff scale from which this suppression occurs (as this parameter sets the free-streaming scale $k_{\rm fs}$). On can see that the suppression of the galaxy power spectrum increases as $\tau$ decreases (as for the previous model), while the suppression starts to occur on larger scales as $\varepsilon$ increases.
The main difference with the previous model is that the suppression of the power spectrum is no longer due to background effects, but to perturbation effects.
At the background level, since one of the daughter particles is massive, the amount of matter can remain roughly equivalent to that of $\Lambda$CDM, which allows the model to accommodate $\Omega_m$ measurements.
At the perturbation level, the presence of the WDM component, which does not cluster on small scales ({\it i.e}, when $k>k_{\rm fs}$), suppresses the galaxy power spectrum.

\subsection{EFTofLSS Constraints on the DCDM $\to$ WDM+DR model}

In the right panel of Fig.~\ref{fig:constraints}, we display the 1D and 2D posteriors of the reconstructed parameters for the DCDM $\to$ WDM+DR model with and without the EFTofBOSS dataset, always including a $S_8$ prior from the KIDS-1000 cosmic shear measurement ($S_8=0.759^{+0.024}_{-0.021}$). Interestingly, we can see that this model has the ability to resolve the $S_8$ tension, and that the inclusion of the EFTofBOSS data does not change this conclusion.
With the EFTofBOSS data, we find $\Delta \chi^2 = -3.8$ with respect to the analogous $\Lambda$CDM analysis (for 2 extra degrees of freedom) at virtually no cost in $\chi^2$ for other data. In particular, the inclusion of the $S_8$ prior help in opening up the degeneracy with the DCDM parameters, without degrading the fit to the other data.~\cite{FrancoAbellan:2021sxk}

In addition, one can see in the right panel of Fig.~\ref{fig:constraints} that the main impact of EFTofBOSS data is to cut in the $\log_{10}(\Gamma/{\rm  Gyr}^{-1})-\log_{10}(\varepsilon)$ degeneracy, excluding too large values of $\log_{10}(\Gamma/{\rm Gyr}^{-1})$.
Therefore, the EFTofLSS significantly improves the constraints on the $\tau= \Gamma^{-1}$ parameter at $1\sigma$: $1.61 < \log_{10}(\tau /{\rm Gyr}) < 3.71$,
to be compared with $1.31 < \rm{log}_{10}(\tau /{\rm Gyr}) < 3.82$ without the EFTofBOSS data.~\cite{Simon:2022ftd} 
Additionally, we observe a notable evolution of the DCDM parameters of the best-fit model compared to the analysis without EFTofBOSS: the best-fit model, with the inclusion of the $S_8$ likelihood, now has $\Gamma = 0.0083 \ \rm{Gyr}^{-1}$ ($\tau = 120 $ Gyr) and $\varepsilon = 0.012 $, while previously $\Gamma = 0.023 \ \rm{Gyr}^{-1}$ ($\tau = 43$ Gyr) and $\varepsilon = 0.006$.
This means that EFTofBOSS data favors longer lived DM models and therefore a smaller fraction of WDM today $f_{\rm wdm}\equiv \bar{\rho}_{\rm wdm}/(\bar{\rho}_{\rm dcdm}+\bar{\rho}_{\rm wdm})\simeq 10 \%$ compared to $f_{\rm wdm}\simeq 27 \%$ previously.

\section*{Acknowledgments}

I am grateful to P. Zhang and G. F. Abellán for their contributions and their kindness.
I am also thankful to V. Poulin, my PhD supervisor, for his benevolence and his extremely wise advice.

\section*{References}

\bibliography{biblio}

\end{document}